# Type II superconductivity in SrPd$_2$Ge$_2$

**T Samuely[1,2], P Szabó[1], Z Pribulová[1], N H Sung[3], B K Cho[3,4], T Klein[5], V Cambel[6], J G Rodrigo[7] and P Samuely[1]**

[1]Centre of Low Temperature Physics, Institute of Experimental Physics Slovak Academy of Sciences & Faculty of Science, P.J. Šafárik University, Košice, Slovakia
[2]INPAC - Institute for Nanoscale Physics and Chemistry, Nanoscale Superconductivity and Magnetism Group, K.U.Leuven, Belgium
[3]School of Materials Science and Engineering, Gwangju Institute of Science and Technology, Gwangju 500-712, Korea
[4]Department of Nanobio Materials and Electronics, Gwangju Institute of Science and Technology, Gwangju 500-712, Korea
[5]Institut Néel, CNRS and Université Joseph Fourier, Boîte Postale 166, 38042 Grenoble, France
[6]Institute of Electrical Engineering, Slovak Academy of Sciences, Bratislava, Slovakia
[7]Laboratorio de Bajas Temperaturas, Depto. de Física de la Materia Condensada and Instituto Nicolas Cabrera, Univesidad Autonoma de Madrid, Spain

E-mail: tomas.samuely@fys.kuleuven.be

**Abstract.** Previous investigations have shown that SrPd$_2$Ge$_2$, a compound isostructural with "122" iron pnictides but iron- and pnictogen-free, is a conventional superconductor with a single *s*-wave energy gap and a strongly three-dimensional electronic structure. In this work we reveal the Abrikosov vortex lattice formed in SrPd$_2$Ge$_2$ when exposed to magnetic field by means of scanning tunneling microscopy and spectroscopy.
Moreover, by examining the differential conductance spectra across a vortex and estimating the upper and lower critical magnetic fields by tunneling spectroscopy and local magnetization measurements, we show that SrPd$_2$Ge$_2$ is a strong type II superconductor with $\kappa \gg 2^{-1/2}$. Also, we compare the differential conductance spectra in various magnetic fields to the pair breaking model of Maki – de Gennes for dirty limit type II superconductor in the gapless region. This way we demonstrate that the type II superconductivity is induced by the sample being in the dirty limit, while in the clean limit it would be a type I superconductor with $\kappa \ll 2^{-1/2}$, in concordance with our previous study (T. Kim et al., Phys. Rev. B **85**, (2012)).

## 1. Introduction

Compounds with the ThCr$_2$Si$_2$ type structure have been extensively studied for decades [1–3]. The discovery of superconductivity in iron pnictides by Hosono et al. [4] and more specifically in pnictide compounds AETM$_2$Pn$_2$ (AE=alkaline earth, TM = transition metal, and Pn = pnictogen) [5,6] boosted the interest in the ThCr$_2$Si$_2$ type structure compounds. This so called "122" stoichiometric family of pnictides features a remarkably high superconducting transition temperature ($T_c \sim 38$ K) when doped with holes.

In general, these iron pnictides are anisotropic, quasi-two-dimensional systems with a crystal structure formed by negatively charged blocks [TM$_2$Pn$_2$]$^{\delta-}$ alternating with positively charged [AE]$^{\delta+}$ blocks. The [TM$_2$Pn$_2$]$^{\delta-}$ blocks account for the superconductivity whilst the [AE]$^{\delta+}$ blocks act as charge reservoirs [7–13]. Still, the mechanism responsible for superconductivity in iron pnictides is unclear to date and the roles of magnetic, chemical and structural properties of the compounds remain an issue. To unravel it, already several pnictide superconductors without magnetic elements were studied [14–19]. These isostructural materials serve as a playground for investigating the role of the structure in the mechanism of superconductivity in the absence of a magnetic element. Regarding these intents, the recently discovered low-temperature superconducting compound SrPd$_2$Ge$_2$ [20] isostructural with the "122" family is even pnictogen-free, in addition to the lack of a magnetic element, and therefore enables a different aspect of the investigation of the superconductivity in "122" iron pnictides.

Shein and Ivanovskii have compared the structural and electronic properties of SrPd$_2$Ge$_2$ and SrNi$_2$As$_2$ [21]. From the band structure calculations they have concluded, that the higher superconducting transition temperature for SrPd$_2$Ge$_2$ in comparison to SrNi$_2$As$_2$ cannot be explained in terms of the electronic factor within the conventional electron – phonon BCS theory, as SrPd$_2$Ge$_2$ has a lower electronic density of states (DOS). Instead, it might be connected to the softening in phonon modes and pronounced electron - phonon coupling induced by the variation of lattice parameters.

In our latter work [22], we have shown, by means of angle resolved photoemission spectroscopy (ARPES), scanning tunneling spectroscopy (STS) and band structure calculations, that SrPd$_2$Ge$_2$, unlike the "122" family of pnictides, has a strongly three-dimensional electronic structure. It is well described within local density approximation (LDA) and features a single isotropic superconducting energy gap with $2\Delta/k_BT_c$ close to the Bardeen – Cooper –Schrieffer (BCS) theory universal value, ruling out exotic electronic states. Yet, in spite of a thorough analysis, the question whether SrPd$_2$Ge$_2$ is a type I or a type II superconductor remained unanswered and is discussed in the text below.

## 2. Experimental methods

The scanning tunneling microscopy (STM) and spectroscopy experiments were performed by means of a homemade STM head in Košice developed in collaboration with UAM Madrid [23], inserted in a commercial Janis SSV cryomagnetic system with $^3$He refrigerator and controlled by Nanotec's Dulcinea SPM electronics.

Single crystals of SrPd$_2$Ge$_2$ were grown by high temperature flux method using PdGe self flux [24]. In order to obtain a clean sample surface, the SrPd$_2$Ge$_2$ sample was cleaved shortly before inserting into the refrigerator and the system was pumped to high vacuum to minimize the contamination of the sample surface. The duration of the whole process was about 15 min. Prior to measurement, the Au tip was prepared in-situ by repetitive impaling into the bulk Au sample and subsequent slow retraction, while recording the current as a function of the tip position at a constant bias voltage. The procedure was repeated until the current – position dependence exhibited clear steps indicating the conductance quantization and single atom contact phenomena typical for gold [25]. The tip was then scanned over the SrPd$_2$Ge$_2$ sample. Bias voltage was applied to the tip, while the sample was grounded; the initial tunneling resistance was set to 1 MΩ. Magnetic field was applied perpendicular to the *ab* plane of the SrPd$_2$Ge$_2$ crystal via a superconducting coil installed in the cryostat.

The magnetization experiments were performed by means of a sensor comprising an array of several rigid Hall probes based on semiconductor heterostructures with a two-dimensional electron gas as the active layer. The array, prepared in Bratislava, consists of 10 Hall probes arranged in a line; the size of the individual probe is 10 x 10 μm$^2$ with the spacing between two neighboring probes 25 μm. The local magnetization of the sample was measured by placing the sample on top of the sensor mounted inside a home-made $^3$He cryostat in Grenoble; the magnetic field was applied perpendicular to the *ab* plane of the SrPd$_2$Ge$_2$ crystal via a superconducting coil installed in the cryostat. The sensor recorded local magnetic induction as the sample was exposed to an external field. The sensor was supplied with constant bias current and the voltage measured across the sensor was directly proportional to the magnetic field penetrating the sample.

## 3. Results and discussion

The differential conductance versus voltage spectrum obtained by STS reveals the convolution of the local DOS of both electrodes comprising the tunneling junction. Figure 1 shows the normalized tunneling conductance spectra between the Au tip and the SrPd$_2$Ge$_2$ sample measured at different temperatures ranging from 0.4 to 2.6 K, increased stepwise by 0.1 K. Each curve was normalized to the spectrum measured for the sample in the normal state. Since the Au tip features a constant density of states, each of these differential conductance versus voltage spectra reflects the superconducting density of states (SDOS) of the SrPd$_2$Ge$_2$, smeared by ~ ± $2k_BT$ in energy at the respective temperature. Consequently, in the low temperature limit ($k_BT \ll \Delta$), the differential conductance measures directly the SDOS. [26]

**Type II superconductivity in SrPd$_2$Ge$_2$**

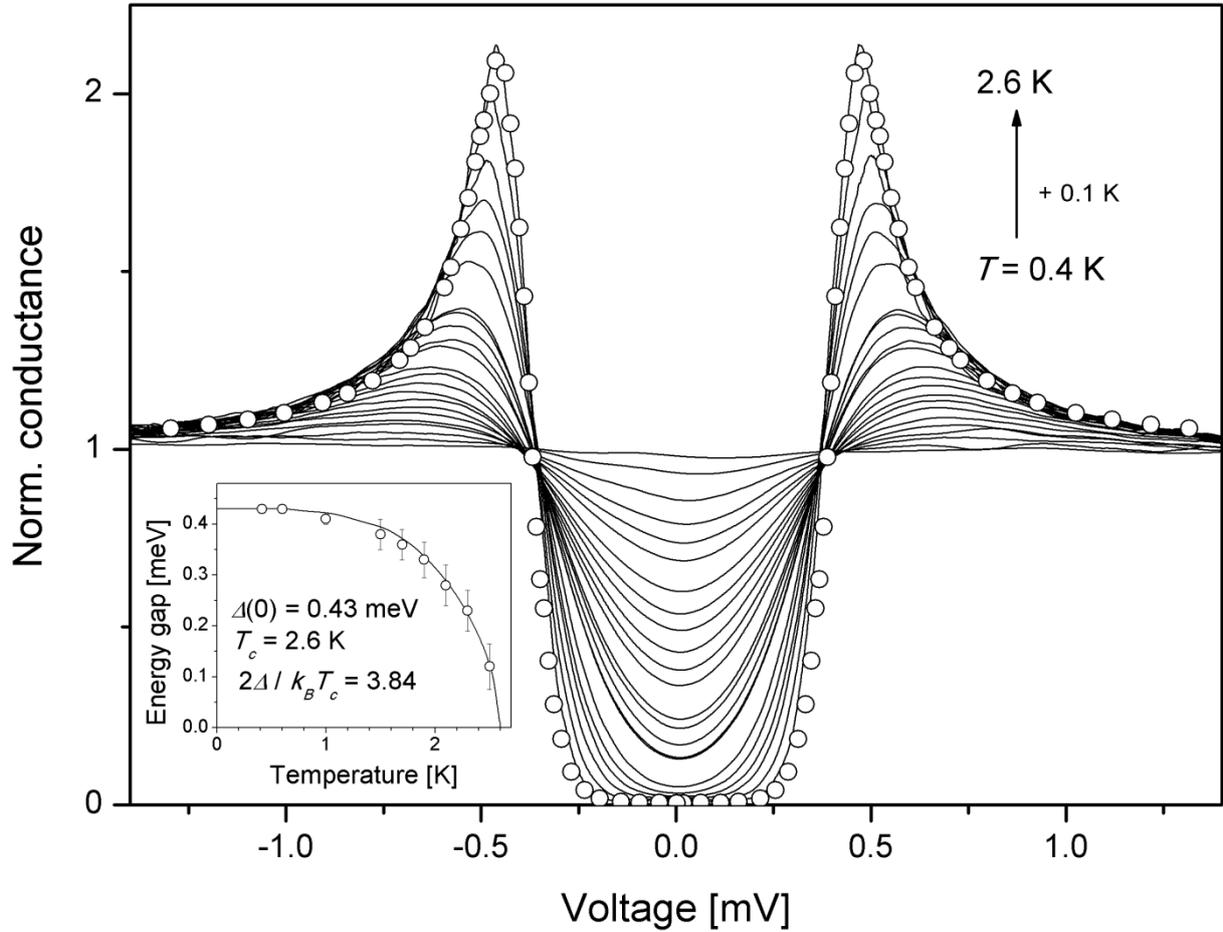

**Figure 1.** Normalized differential conductance spectra acquired by STS, measured between the Au tip and the SrPd$_2$Ge$_2$ sample in zero magnetic field at different temperatures between 0.4 K and 2.6 K, increasing by 0.1 K. The BCS fit of the spectrum obtained at the lowest temperature ($T$ = 0.4 K) is indicated by circles. The inset shows the temperature dependence of the superconducting gap of SrPd$_2$Ge$_2$ (circles), obtained by the BCS fit of individual curves, in comparison with the BCS theory (line).

By fitting the measured spectra to the BCS density of states $N(E) = Re\{E/(E^2-\Delta^2)^{1/2}\}$, where $E$ is the quasiparticle energy and $\Delta$ the superconducting energy gap of the SrPd$_2$Ge$_2$, convoluted with the Fermi distribution function, the temperature dependence of the superconducting gap of SrPd$_2$Ge$_2$ was obtained. Apart from the thermal smearing, no other smearing parameter was employed. As shown in the inset of Figure 1, temperature dependence of the superconducting gap of SrPd$_2$Ge$_2$ (circles) coincides accurately with the prediction of the BCS theory (line), yielding the superconducting energy gap value $\Delta(0) = 0.43$ meV and the critical temperature $T_c = 2.6$ K. This indicates medium coupling superconductivity with a ratio of $2\Delta/k_BT_c = 3.84$, enhanced by *ca.* 10% compared to the BCS theory universal value. The minute discrepancies between the above given values of $\Delta(0)$ and $T_c$ and the ones presented in the previous study [22] possibly originate from the variations occurring in the sample fabrication process, just as different values of $T_c$ were observed for a polycrystalline sample [20] and a single crystal [24].

From ARPES, Kim et al. determined the Fermi velocity $\hbar v_F = 4.7$ eVÅ and the penetration depth $\lambda_0 = 40$ nm for SrPd$_2$Ge$_2$ [22]. In combination with the above given superconducting energy gap value $\Delta(0)$, the coherence length $\xi_0 = 348$ nm was estimated from

$$\xi_0 = \frac{\hbar v_F}{\pi\Delta(0)}, \qquad (1)$$

providing the Ginzburg – Landau parameter $\kappa_0 = \lambda_0/\xi_0 = 0.11$, which implies that SrPd$_2$Ge$_2$ is likely to be a type-I superconductor. Contrarily, as described in the following text, the behavior of SrPd$_2$Ge$_2$ in magnetic field proved otherwise and confirmed the preliminary indirect indications of type II superconductivity introduced in the former study [22].

The influence of the magnetic field on the differential conductance between the Au tip and the SrPd$_2$Ge$_2$ sample is shown in Figure 2 (a). The differential conductance spectra were measured in magnetic fields ranging from zero up to 500 mT in steps of 20 mT at 0.42 K. Strikingly, a non-

**Type II superconductivity in SrPd$_2$Ge$_2$**

monotone behavior of the zero bias conductance (ZBC) as a function of the applied magnetic field $B$ was observed, as exposed in the inset of Figure 2 (a). This behavior can be explained by the dynamics of superconducting vortices, magnetic flux quanta penetrating the SrPd$_2$Ge$_2$ sample, which locally disrupt superconductivity and suppress the superconducting order parameter. In analogy to the "Lazy Fisherman" method introduced by Kohen et al. [27], by changing the value of the applied magnetic field, vortex motion below the stationary Au tip was induced. As a result, different points within the vortex lattice were accessed, leading to an increase of the ZBC in the vicinity of a vortex core and a decrease of the ZBC far from the vortex core.

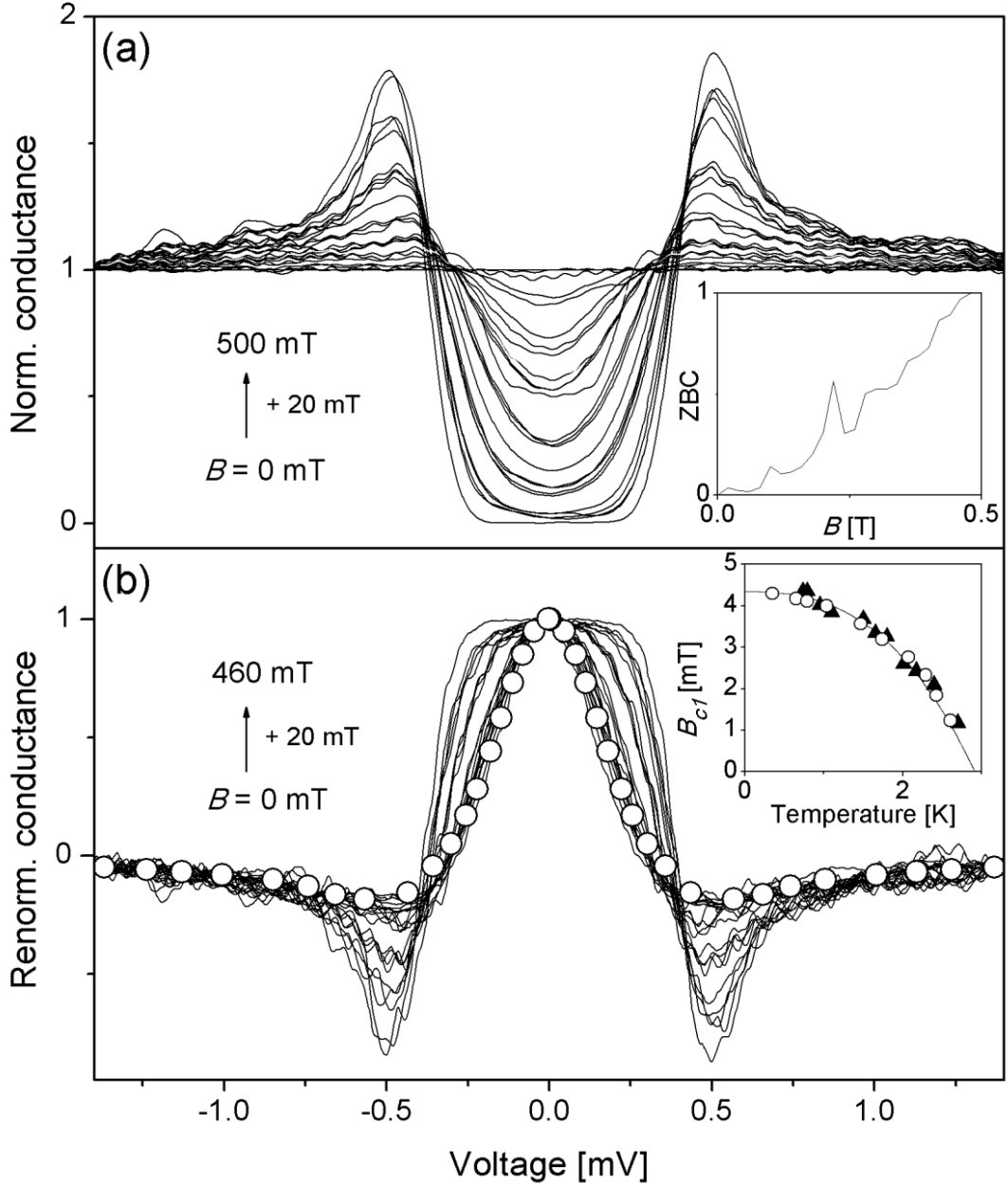

**Figure 2.** (a) Normalized differential conductance spectra acquired by STS, measured between the Au tip and the SrPd$_2$Ge$_2$ sample at 0.4 K in different magnetic fields between 0 mT and 500 mT, increasing by 0.02 T. Inset of (a): Normalized ZBC as a function of the applied magnetic field. (b) Differential conductance spectra renormalized to their zero bias value according to $[N(E) - N_N(0)]/[N(0) - N_N(0)]$ in magnetic fields between 0 mT and 460 mT. The renormalized de Gennes formula (6) of the SDOS fitted to the curves is indicated by open circles, yielding the pair-breaking parameter $\alpha = 0.32 \pm 0.03$ meV. The differential conductance spectra in magnetic fields above 460 mT are excluded from the figure due to the excessive noise amplification stemming from the renormalization. Inset of (b): Temperature dependence of $B_{c1}$ measured at two different Hall probes sensor positions, yielding $B_{c1} = 4.35 \pm 0.5$ mT in the limit of $T = 0$ K.

# Type II superconductivity in SrPd$_2$Ge$_2$

Besides, the estimated upper critical magnetic field $B_{c2}$ = 500 ± 20 mT corresponds well to the value obtained by Sung et al. from magnetization and heat capacity measurements [24] with regard to the Werthamer – Helfand – Hohenberg theory [28]. The effective coherence length $\xi$ was calculated from the low-temperature upper-critical-field value $B_{c2}$ as

$$\xi = \sqrt{\frac{\Phi_0}{2\pi B_{c2}}}, \tag{2}$$

where $\phi_0$ is the flux quantum, giving $\xi$ = 25.6 ± 0.5 nm, a value considerably smaller than $\xi_0$.

Furthermore, the local magnetization measurements of SrPd$_2$Ge$_2$ unveiled the penetration of magnetic field lower than $B_{c2}$ and enabled the estimation of the lower critical magnetic field $B_{c1}$. At magnetic fields lower than $B_{c1}$, the sample was in the Meissner state expelling the field from its interior and distorting its distribution in the vicinity of the sample surface. Due to this distortion, the sample was exposed to an effectively larger field compared to the applied one. The first vortex penetrated the sample when this effective field reached the value of $B_{c1}$ all around the sample surface, while the applied field read the value $B_p < B_{c1}$, $B_p$ being the first penetration field. In samples with arbitrary cross sections a demagnetization coefficient needs to be considered in order to determine the lower critical magnetic field $B_{c1}$ [29]. In the presence of geometrical barriers $B_p$ is related to $B_{c1}$ through

$$\frac{B_p}{B_{c1}} \sim \tanh\sqrt{\frac{\sigma d}{w}}, \tag{3}$$

where $\sigma$ is a constant varying from 0.36 in strips to 0.67 in disks, $w$ is the sample width and $d$ is the sample thickness, with a ratio $d/w \sim 0.3$ in our case. For fields lower than the first penetration field $B_p$ no magnetic field penetrated the sample, thus the sensor was shielded and no voltage was detected. When the field increased above $B_p$, the local magnetization sensor started to read increasing voltage as the number of vortices grew. The local magnetization measurements were performed at various temperatures for several different positions of the sensor with respect to the sample edge, as described more in detail by Rodière et al. [30] The temperature dependence of $B_{c1}$ measured at two different sensor positions is displayed in the inset of Figure 2 (b). As shown, $B_{c1}$ clearly flattens off at low temperatures indicating that the superconducting energy gap was fully open. In the limit of $T$ = 0 K we acquired the value of $B_{c1}$ = 4.3 ± 0.5 mT. Exploiting the following expressions relating $B_{c1}$ and $B_{c2}$ derived by Brandt [31]:

$$\frac{B_{c1}}{B_{c2}} = \frac{\ln\kappa + \delta(\kappa)}{2\kappa^2}; \delta(\kappa) = 0.5 + \frac{1+\ln 2}{2\kappa - \sqrt{2} + 2} \tag{4}$$

permitted us to gain the effective Ginzburg – Landau parameter $\kappa$ = 13.5 ± 1 along with the effective penetration depth $\lambda = \kappa\xi$ = 345 ± 30 nm, much larger than $\lambda_0$. The value of $\kappa$ is in accordance with the expected type II superconductivity in SrPd$_2$Ge$_2$ and in steep contrast to $\kappa_0$ = 0.11 estimated from the penetration depth $\lambda_0$ and the coherence length $\xi_0$.

Ultimately, the above mentioned indications of type II superconductivity in SrPd$_2$Ge$_2$ were confirmed by the Conductance Imaging Tunneling Spectroscopy (CITS) measurements [32] performed at 0.42 K in various magnetic fields, disclosing the presence of superconducting vortices. In these experiments, the scanned surface of 500 × 500 nm$^2$ was divided into 128 x 128 points and in each point a differential conductance spectrum was recorded. Subsequently, the conductance value at 0.119 mV was plotted. Conductance plots at other voltage values within the superconducting gap were similar in appearance. The interpretation of these plotted differential conductance maps is following:

Since 0.119 meV < $\Delta$, no tunneling current can flow in the superconducting region as opposed to the regions where the superconducting vortices penetrate the sample. Inside the vortex region, the superconducting energy gap is suppressed gradually from the periphery of the vortex core towards its center, where the superconductivity is entirely disrupted. Hence, the dark regions correspond to superconducting state and the bright blots to the regions with suppressed superconductivity.

Figures 3 (a), (b) and (c) show the differential conductance maps recorded at magnetic fields of 50 mT, 100 mT and 250 mT, respectively. Even though the bright gapless regions do not appear as regularly shaped vortices, the variation of the order parameter across the scanned surface is evident. The blurry appearance of the interfaces between fully gapped and gapless regions is caused by the ample corrugation of the sample surface that considerably obstructs fine resolved STM imaging. Figure 3 (d) shows the topography of the scanned area without the applied magnetic field. The roughness of the surface is demonstrated by the cross section at the bottom of the image.

**Type II superconductivity in SrPd$_2$Ge$_2$**

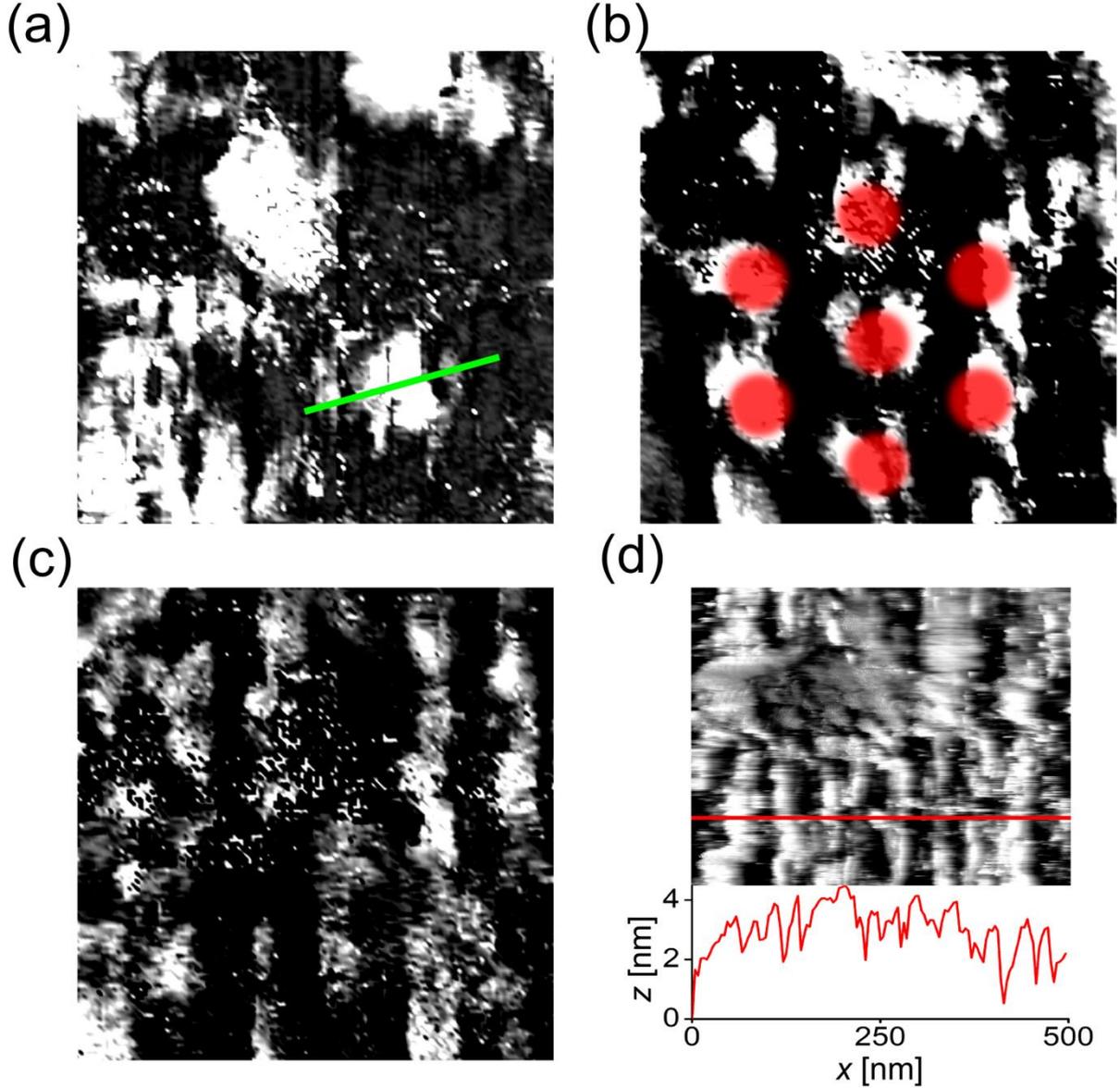

**Figure 3.** (a), (b) and (c): Conductance imaging tunneling spectroscopy (CITS) maps (500 × 500 nm$^2$, 0.119 mV, 0.42 K) of the SrPd$_2$Ge$_2$ surface in magnetic fields of 50 mT (a), 100 mT (b) and 250 mT (c), illustrating the superficial variation of the sample DOS at 0.119 meV. Red transparent circles in (b) represent the Abrikosov lattice. The diameter of each circle is $2\xi$ and they are separated by $a = 155$ nm, the Abrikosov lattice constant for a magnetic field of 100 mT. (d) Top: STM topographic image (500 × 500 nm$^2$, 1 nA, 1 mV, 0.42 K) of the area shown in (a), (b) and (c) without the applied magnetic field. Bottom: The line profile of the topography along the red line in the STM image above.

Moreover, the density of the gapless regions represented by the bright blots increases with the increase of the applied magnetic field and the order parameter varies across the scanned surface with periodicity roughly matching the Abrikosov lattice constant

$$a = \left(\frac{4}{3}\right)^{\frac{1}{4}} \sqrt{\left(\frac{\Phi_0}{B}\right)} \qquad (5)$$

for the corresponding magnetic field $B$, i.e. 219 nm for 50 mT, 155 nm for 100 mT and 98 nm for 250 mT. This is most easily discerned in Figure 3 (b) recorded at 100 mT, where a tentative model of the Abrikosov vortex lattice represented by red transparent circles is superimposed. The diameter of each circle is $2\xi$ and the distance between two adjacent circles is $a = 155$ nm.

Because in each point of the CITS measurement the entire differential conductance spectrum was recorded, the evolution of the SDOS across a vortex can be visualized. Figure 4 (a) shows the normalized differential conductance spectra along a 216 nm line across a bright blot in $B = 50$ mT. The location is indicated by the green line in Figure 3 (a). The difference between the normal DOS at

**Type II superconductivity in SrPd$_2$Ge$_2$**

the vortex center and the SDOS far from it is obvious and the order parameter is suppressed over a length scale of 2$\xi$ [33]. Also, at the vortex core, the differential conductance spectrum is featureless; no ZBC peak induced by the localized quasiparticle states is present. This suggests that the sample is in the dirty limit, implying a rather short electron mean free path $l$ [34].

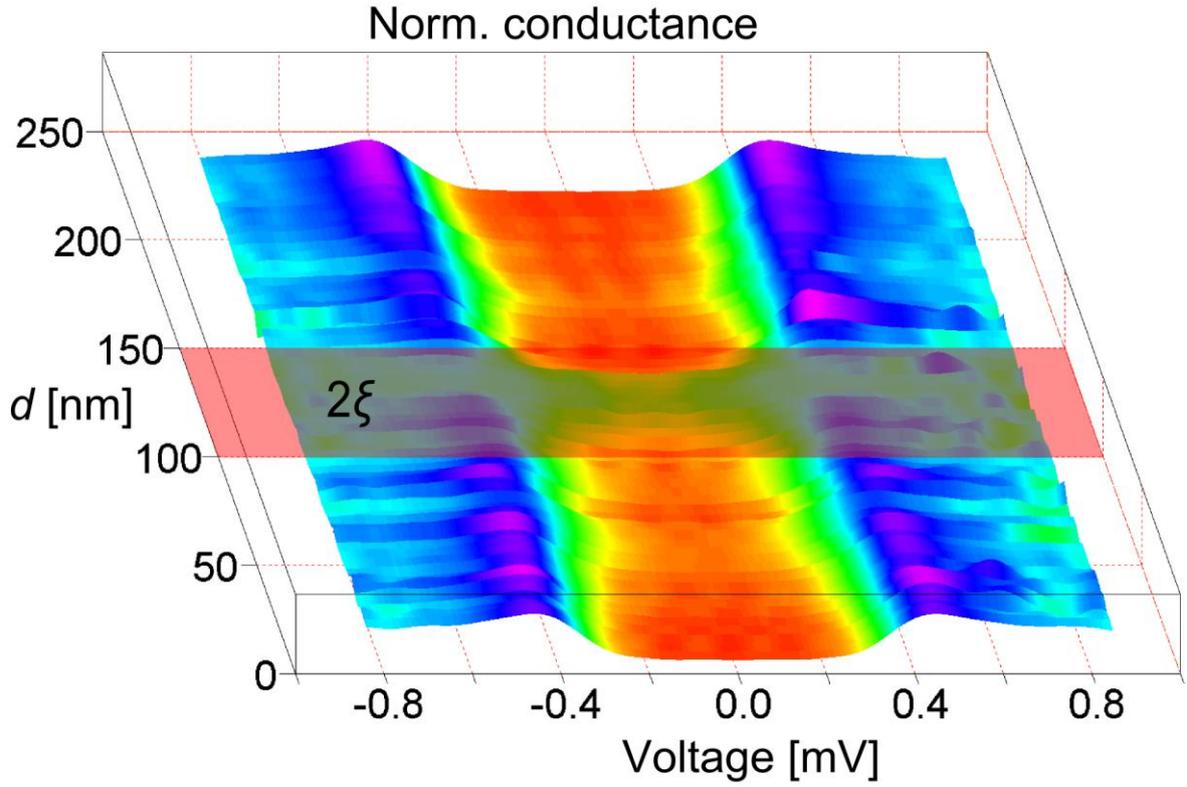

**Figure 4.** Normalized differential conductance spectra along the green line in Figure 3 (a). Length of the line is 216 nm. The spectra demonstrate the variation of the SDOS across a vortex in the applied magnetic field of 50 mT. The vortex center is at $d \sim 125$ nm, featuring a normal DOS. Far from the vortex center, a full superconducting gap is present. The red transparent stripe marks the distance 2$\xi$, roughly matching the vortex core size.

The electron mean free path was not considered in the estimation of $\lambda_0$ and $\xi_0$ using only the Fermi velocity, superconducting energy gap and the averaged length of the Fermi contours [22] and concurs with the fact that the effective penetration depth $\lambda$ is much larger than $\lambda_0$ and the effective coherence length $\xi$ is much smaller than $\xi_0$.

The notion of the dirty limit is further validated by the applicability of the pair breaking model of de Gennes [35] and Maki [36] to the differential conductance spectra in various magnetic fields (Figure 2 (a)), as described by Szabó et al. [37] In this model, the SDOS $N(E)$ of a dirty limit type II superconductor in the gapless region for small $\Delta$ in magnetic fields near $B_{c2}$ is given by

$$N(E) = N_N(0)\left[1 - \frac{\Delta^2(B)}{2}\frac{\alpha^2 - E^2}{(\alpha^2 + E^2)^2}\right], \tag{6}$$

where $N_N(0)$ is the density of states at the Fermi surface of the superconductor in the normal state and $\alpha$ is the Maki – de Gennes pair-breaking parameter.

From the model follows, that after renormalizing the SDOS using the expression $[N(E) - N_N(0)]/[N(0) - N_N(0)]$, the field-dependent parameter $\Delta$ falls out of the problem. As stated above, the differential tunneling conductance is directly proportional to the SDOS. As a consequence, all conductance curves, renormalized to their zero bias value according to the latter expression, collapse onto a single curve above a certain magnetic field ($\sim 0.5B_{c2}$ in our case). This is demonstrated in Figure 2 (b). By fitting the expression (6) in its renormalized form to the resultant curve, the pair-breaking parameter $\alpha = 0.32 \pm 0.03$ meV was gained directly. For weak-coupling superconductors in the low temperature limit, the pair-breaking coefficient relates to the energy gap as $\alpha = \Delta(0)/2$, whereas in our case, the value is about 50 % higher. Such divergence was observed also for other conventional type II superconductors in the dirty limit and with stronger coupling. [38–40] Additionally, the pair-breaking parameter $\alpha$ relates the electron mean free path $l$ to the effective coherence length $\xi$. Therefore it was possible to calculate the mean free path using the relation

Type II superconductivity in SrPd$_2$Ge$_2$

$$l = \frac{6\alpha\xi^2}{\hbar v_F}. \quad (7)$$

Considering the above given values of the pair-breaking parameter $\alpha$, Fermi velocity $v_F$, and the effective coherence length $\xi$, we obtained the value of the electron mean free path $l = 2.7 \pm 0.3$ nm.

Using the above estimated value of the electron mean free path $l$ and the experimentally confirmed effective coherence length $\xi$ and effective penetration depth $\lambda$, we calculated the clean limit values of the Pippard coherence length $\xi_P$ and the London penetration depth $\lambda_L$ in the context of Ginzburg – Landau theory from the relations

$$\xi_P = \frac{\xi^2}{0.731 l}; \lambda_L = \lambda\sqrt{\frac{1.33 l}{\xi_P}}. \quad (8)$$

The calculated values $\xi_P = 332 \pm 45$ nm and $\lambda_L = 36 \pm 5$ nm match the values $\xi_0$ and $\lambda_0$ introduced above. These values are valid for the sample in the clean limit, not considering the mean free path of electrons. On the contrary, the effective values $\xi$ and $\lambda$ obtained in this study are valid for a sample in the dirty limit, with the electron mean free path $l$ substantially smaller than the Pippard coherence length $\xi_P$.

## 4. Conclusion

To summarize, the differential conductance spectra gained from tunneling between the normal metal tip (Au) and the superconducting sample (SrPd$_2$Ge$_2$) at various temperatures have confirmed the single s-wave character and BCS-like temperature dependence of the medium size gap of SrPd$_2$Ge$_2$.

What is more, by means of STM/S we have visualized the magnetic field penetrating the sample in the form of superconducting vortices and estimated the effective Ginzburg – Landau parameter $\kappa = 13.5$ by analyzing the critical magnetic fields. Hence, we have shown strong type II superconductivity in SrPd$_2$Ge$_2$, confirming the preliminary indirect indications presented in our previous study [22] and the magnetization studies [20,24].

Finally, we have addressed the ostensible discrepancy between our estimate of $\kappa$ and the significantly lower value $\kappa_0 = 0.11$ obtained from ARPES and STS data [22], indicating that SrPd$_2$Ge$_2$ is likely to be a type I superconductor. A viable explanation of this is that the type II superconductivity is induced by the sample being in dirty limit. We have validated this by comparing the behavior of the SDOS of the sample in various magnetic fields to the Maki and de Gennes [35,36] model of gapless superconductivity, showing that it is consistent with the behavior of a superconductor in a dirty limit. Moreover, by estimating the electron mean free path, we have been able to identify the Pippard coherence length and the London penetration depth valid in the clean limit with the values of the coherence length and penetration depth obtained from ARPES and STS data [22], which indicate type I superconductivity. In view of that, while the type I superconductivity has been attributed to the clean limit, in the dirty limit SrPd$_2$Ge$_2$ behaves as a type II superconductor.

**Acknowledgements**
This work has been supported by ESF Research Networking Program Nanoscience and Engineering in Superconductivity NES, 7th FP MNT – ERA.Net II. ESO, the Slovak Research and Development Agency under the contract No. 0036 - 11, the Slovak Grant VEGA Nos. 2/0148/10 and 1/0782/12. The research leading to these results has received funding from the European Community`s Seventh Framework Programme (FP7/2007-2013) under grant agreement n° 228464 (MICROKELVIN) and the project 2011-0028736 of the Ministry of Education, Science and Technology of Korea. The liquid nitrogen for the experiment has been sponsored by the US Steel Kosice, s.r.o. The Center of Low Temperature Physics is operated as the Center of Excellence CFNT MVEP of the Slovak Academy of Sciences. JGR acknowledges support from projects FIS2011-23488 and Consolider-Ingenio Molecular Nanoscience CSD2007-00010.

**References**
[1] Mathur N D, Grosche F M, Julian S R, Walker I R, Freye D M, Haselwimmer R K W and Lonzarich G G 1998 *Nature* **394** 39–43
[2] Trovarelli O, Geibel C, Mederle S, Langhammer C, Grosche F, Gegenwart P, Lang M, Sparn G and Steglich F 2000 *Phys. Rev. Lett.* **85** 626–9
[3] Gegenwart P, Westerkamp T, Krellner C, Tokiwa Y, Paschen S, Geibel C, Steglich F, Abrahams E and Si Q 2007 *Science (New York, N.Y.)* **315** 969–71
[4] Kamihara Y, Watanabe T, Hirano M and Hosono H 2008 *J. Am. Chem. Soc.* **130** 3296–7
[5] Rotter M, Tegel M and Johrendt D 2008 *Phys. Rev. Lett.* **101** 107006

# Type II superconductivity in SrPd$_2$Ge$_2$

[6]  Sasmal K, Lv B, Lorenz B, Guloy A M, Chen F, Xue Y-Y and Chu C-W 2008 *Phys. Rev. Lett.* **101** 107007
[7]  Sadovskii M V 2008 *Physics-Uspekhi* **51** 1201–27
[8]  Ivanovskii A L 2008 *Physics-Uspekhi* **51** 1229–60
[9]  Izyumov Y A and Kurmaev E Z 2008 *Physics-Uspekhi* **51** 1261–86
[10] Rotter M, Tegel M and Johrendt D 2008 *Phys. Rev. B* **78** 20503
[11] Zhu X, Han F, Mu G, Zeng B, Cheng P, Shen B and Wen H-H 2009 *Phys. Rev. B* **79** 24516
[12] Hsu F-C, Luo J-Y, Yeh K-W, Chen T-K, Huang T-W, Wu P M, Lee Y-C, Huang Y-L, Chu Y-Y, Yan D-C and Wu M-K 2008 *Proc. Natl. Acad. Sci. U. S. A.* **105** 14262–4
[13] Shein I and Ivanovskii A 2009 *Phys. Rev. B* **79** 245115
[14] Jeitschko W, Glaum R and Boonk L 1987 *J. Solid State Chem.* **69** 93–100
[15] Berry N, Capan C, Seyfarth G, Bianchi A D, Ziller J and Fisk Z 2009 *Phys. Rev. B* **79** 180502
[16] Han J-T, Zhou J-S, Cheng J-G and Goodenough J B 2010 *J. Am. Chem. Soc.* **132** 908–9
[17] Shein I and Ivanovskii A 2009 *Phys. Rev. B* **79** 54510
[18] Nath R, Singh Y and Johnston D 2009 *Phys. Rev. B* **79** 174513
[19] Singh Y, Lee Y, Nandi S, Kreyssig A, Ellern A, Das S, Nath R, Harmon B, Goldman A and Johnston D 2008 *Phys. Rev. B* **78** 104512
[20] Fujii H and Sato A 2009 *Phys. Rev. B* **79** 224522
[21] Shein I R and Ivanovskii A L 2010 *Physica B* **405** 3213–6
[22] Kim T, Yaresko A, Zabolotnyy V, Kordyuk A, Evtushinsky D, Sung N, Cho B, Samuely T, Szabó P, Rodrigo J, Park J, Inosov D, Samuely P, Büchner B and Borisenko S 2012 *Phys. Rev. B* **85** 014520
[23] Rodrigo J G, Suderow H, Vieira S, Bascones E and Guinea F 2004 *J. Phys.: Condens. Matter* **16** R1151–R1182
[24] Sung N, Rhyee J-S and Cho B 2011 *Phys. Rev. B* **83** 094511
[25] Agraït N 2003 *Phys. Rep.* **377** 81–279
[26] Tinkham M 2004 *Introduction to Superconductivity* 2$^{nd}$ edn, vol 1 (Mineola, NY: Dover)
[27] Kohen A, Cren T, Proslier T, Noat Y, Sacks W, Roditchev D, Giubileo F, Bobba F, Cucolo A M, Zhigadlo N, Kazakov S M and Karpinski J 2005 *Appl. Phys. Lett.* **86** 212503
[28] Werthamer N R, Helfand E and Hohenberg P C 1966 *Phys. Rev.* **147** 295–302
[29] Brandt E H 1999 *Phys. Rev. B* **59** 3369–72
[30] Rodière P, Klein T, Lemberger L, Hasselbach K, Demuer A, Kačmarčik J, Wang Z, Luo H, Lu X, Wen H, Gucmann F and Marcenat C 2012 *Phys. Rev. B* **85** 214506
[31] Brandt E H 2003 *Phys. Rev. B* **68** 54506
[32] Hamers R, Tromp R and Demuth J 1986 *Phys. Rev. Lett.* **56** 1972–5
[33] Golubov A and Hartmann U 1994 *Phys. Rev. Lett.* **72** 3602–5
[34] Renner C, Kent A, Niedermann P, Fischer Ø and Lévy F 1991 *Phys. Rev. Lett.* **67** 1650–2
[35] Gennes P G 1964 *Phys. Kondens. Mater.* **3** 79–90
[36] Maki K 1964 *Physics* vol 1 (Long Island City, NY) p 21
[37] Szabó P, Samuely P, Jansen A, Marcus J and Wyder P 2000 *Phys. Rev. B* **62** 3502–7
[38] Samuely P, Bobrov N, Jansen A, Wyder P, Barilo S and Shiryaev S 1993 *Phys. Rev. B* **48** 13904–10
[39] Guyon E, Meunier F and Thompson R 1967 *Phys. Rev.* **156** 452–69
[40] Guyon E, Martinet A, Matricon J and Pincus P 1965 *Phys. Rev.* **138** A746–A752